\documentclass[prl,twocolumn, superscriptaddress]{revtex4}
\usepackage{graphicx}
\usepackage{amssymb}
\usepackage{amsmath}
\usepackage{ulem}

\begin{document}
\title{\ A direct quantitative measure of surface mobility in a glassy polymer}
\author{Y. Chai}
\affiliation{Department of Physics \& Astronomy and Guelph-Waterloo Physics Institute, University of Waterloo,  Waterloo, Ontario, Canada, N2L 3G1.}
\author{T. Salez}
\affiliation{Laboratoire de Physico-Chimie Th\'eorique, UMR CNRS Gulliver 7083, ESPCI ParisTech, Paris, France}
\author{J. D. McGraw}
\altaffiliation{Department of Experimental Physics, Saarland University, 66041 Saarbr\"{u}cken, Germany}
\affiliation{Department of Physics and Astronomy, McMaster University,  Hamilton, Ontario, Canada, L8S 4M1.}
\author{M. Benzaquen}
\affiliation{Laboratoire de Physico-Chimie Th\'eorique, UMR CNRS Gulliver 7083, ESPCI ParisTech, Paris, France}
\author{K. Dalnoki-Veress}
\affiliation{Laboratoire de Physico-Chimie Th\'eorique, UMR CNRS Gulliver 7083, ESPCI ParisTech, Paris, France}
\affiliation{Department of Physics and Astronomy, McMaster University,  Hamilton, Ontario, Canada, L8S 4M1.}
\author{E. Rapha\"{e}l}
\affiliation{Laboratoire de Physico-Chimie Th\'eorique, UMR CNRS Gulliver 7083, ESPCI ParisTech, Paris, France}
\author{J. A. Forrest}
\thanks{corresponding author. email: jforrest@uwaterloo.ca}
\affiliation{Department of Physics \& Astronomy and Guelph-Waterloo Physics Institute, University of Waterloo,  Waterloo, Ontario, Canada, N2L 3G1.}
\date{\today}
\begin{abstract}
Thin polymer films have striking dynamical properties that differ from their bulk counterparts. With the simple geometry of a stepped polymer film on a substrate, we probe mobility above and below the glass transition temperature $T_{\textrm{g}}$. Above $T_{\textrm{g}}$ the entire film flows, while below $T_{\textrm{g}}$ only the near surface region responds to the excess interfacial energy. An analytical thin film model for flow limited to the free surface region shows excellent agreement with sub-$T_{\textrm{g}}$ data. The system transitions from whole film flow to surface localized flow over a narrow temperature region near the bulk $T_{\textrm{g}}$. The experiments and model provide a measure of surface mobility in a sample geometry where confinement and substrate effects are negligible. This fine control of the glassy rheology is of key interest to nanolithography among numerous other applications.
\end{abstract}
\maketitle

The last decades have seen a significant interest in the dynamical and rheological properties of glassy materials~\cite{edigerreview,andersonscience}. Recent efforts~\cite{edigerreview,EPJEreview,JAFKDVreview,McKennareview} have focussed on elucidating the nature of glassy dynamics both in the bulk, and in systems such as thin films or colloids where the interfaces play a dominant role and can induce strong dynamical heterogeneities. Higher mobility near interfaces has often been suggested as the cause of anomalous glass transition temperatures in thin polymer films~\cite{EPJEreview,JAFKDVreview,McKennareview}. The presence of a more mobile surface has practical implications for thin film coatings related to lubrication, wear, and friction. Flow on a near surface layer can also place strict lower limits on feasible length scales for nanolithography~\cite{teisseire11APL,rognin11PRE,rognin12JVS}. While earlier investigations provided some contradictory conclusions~\cite{contradict1,contradict2}, most recent reports are consistent with a region of enhanced mobility on the surface of glassy polymer films. There have also been reports of enhanced surface mobility in small molecule glasses~\cite{zhu11PRL,edigerjcp}. The parallels between polymeric and small molecule glasses suggest that enhanced surface mobility is a more general property of glass-forming materials. Most current research efforts have a goal of providing a quantitative description, including the temperature dependence, of the properties of the near surface region. Surface response to nanoparticle embedding has been used to probe anomalous surface dynamics in both polymeric and small molecule glasses~\cite{embeddingprl,embeddingilton,embeddingQi,embeddingMw,ChadTNB}. That work showed that small molecules can flow on the surface, while larger polymers have enhanced segmental mobility, but do not flow owing to their larger molecular size. Relaxation of an imposed surface topography has been used to demonstrate enhanced mobility in polymeric~\cite{zahrascience,QiPRL,Papaleo,buck04MAC} and small molecule systems~\cite{zhu11PRL}. For small molecule glasses, the enhanced surface mobility is often discussed in terms of surface diffusion~\cite{Mullins}, where the molecules at the free surface have a diffusion time that can be orders of magnitude smaller than the bulk value~\cite{zhu11PRL,edigerjcp}. For polymers, the most complete description comes from studies of low molecular weight polystyrene~\cite{tsuiscience}. 
\begin{figure}[t!] 
\includegraphics[width=1\columnwidth]{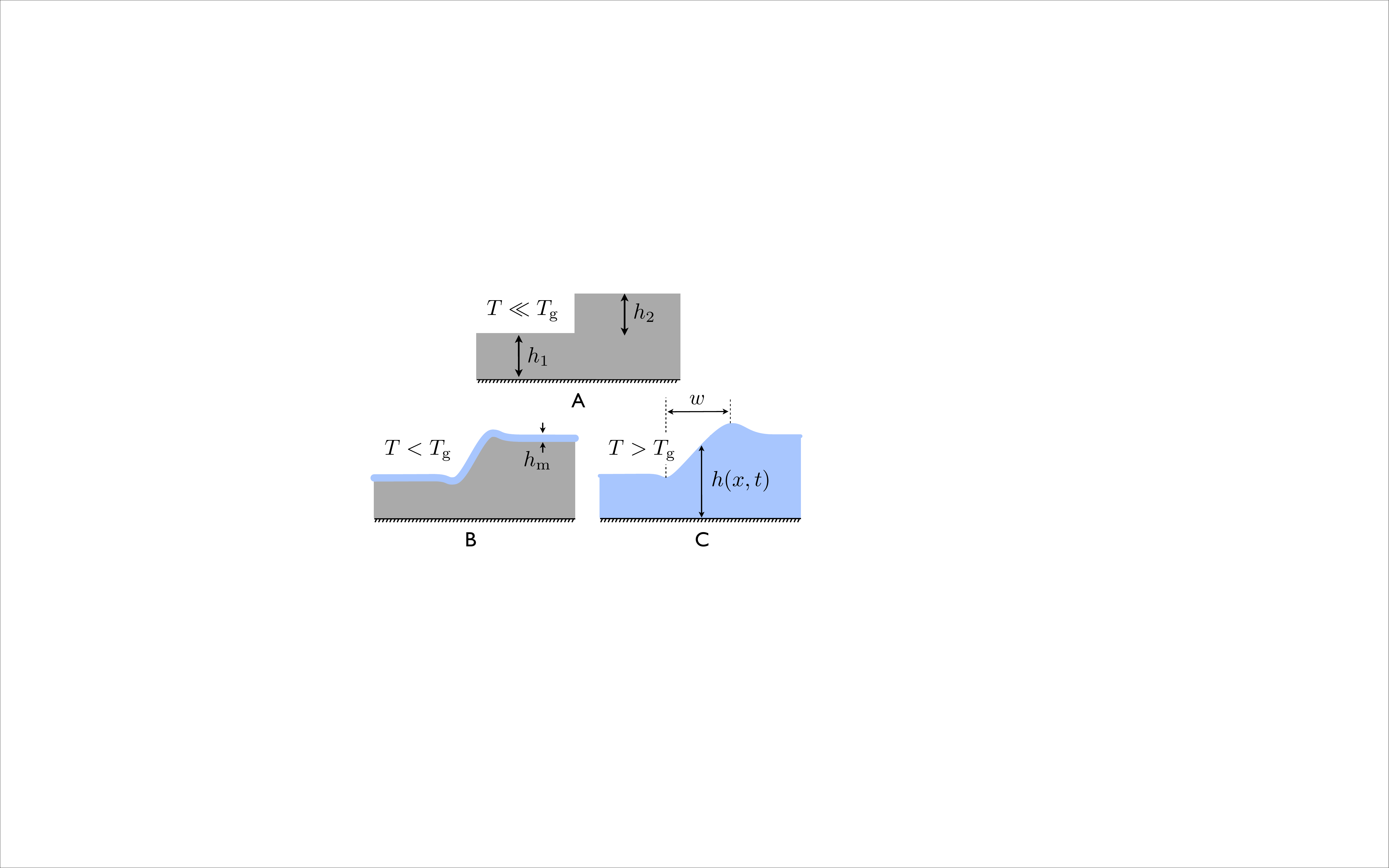}
\caption{Fig. 1. Schematic diagram of the sample geometry and flow regions. \textbf{(A)} An as-prepared sample at room temperature. \textbf{(B} and \textbf{C)} The two flow mechanisms discussed in the paper. \textbf{(C)} describes the evolution of the total height profile $h(x,t)$ through whole film flow (TFE, see Eq.~(\ref{fulltfe})), while \textbf{(B)} shows the evolution of the total height profile through flow localized in a small region near the free surface (GTFE, see Eq.~(\ref{fullgtfe})). The flow region is indicated in blue and is assumed to vanish far below the glass transition temperature $T_{\textrm{g}}$.}
\label{fig1}
\end{figure}
 
The use of capillary leveling as a probe of rheology on the nanometer scale~\cite{McGrawPRL,Salez2012a,Salez2012b} has been successfully used to study polymer rheology for films at temperatures much greater than the $T_{\textrm{g}}$ value of the polymer. Stepped films were annealed, and a decrease in the surface area was monitored to probe dissipation of the system's free energy, with a complete quantification of the rheological properties~\cite{McGrawPRL, Salez2012a, Salez2012b}. The low molecular weight films considered here are sufficiently thin so that gravitational effects can be ignored, yet thick enough so that van der Waals interactions resulting in a disjoining pressure can  be neglected. Gradients in the curvature of the free surface result in Laplace pressure gradients which drive viscous flows. When the height gradients are sufficiently small, and the typical height of the profile is sufficiently smaller than its typical width, the flow can be described by the Stokes equations in the lubrication approximation. For homogenous viscous films, the evolution of the profile $h(x,t)$ is described by the capillary-driven thin film equation (TFE)~\cite{Blossey}: 
\begin{equation}
\label{fulltfe}
\partial_th+\frac{\gamma}{3\eta_{\textrm{b}}}\partial_x(h^3 \partial_x^{\,3}h)=0\ ,
\end{equation} 
where $\eta_{\textrm{b}}$ is the bulk viscosity, $\gamma$ is the surface tension, $x$ is the horizontal coordinate, and $t$ the time. The TFE can be solved numerically for a stepped initial profile~\cite{Salez2012a} and the solution has been shown to converge in time towards a self-similar profile in the variable $xt^{-1/4}$. 

For the case of films with $T<T_{\textrm{g}}$, the majority of the film is unable to flow. Since previous studies have shown an evolution of the free surface  to minimize surface area and energy in polymeric~\cite{zahrascience,QiPRL}  and non-polymeric glasses~\cite{edigerjcp,ChadTNB}, there must be some flow localized over a thin layer near the free surface. At a given temperature, we will assume the thickness $h_{\textrm{m}}$ of this mobile layer, with viscosity $\eta_{\textrm{m}}$, to be constant (see Fig.~1B). Of course, the present two-layer model is an approximation and one would expect a continuous variation from surface to bulk dynamics through the sample~\cite{Forrest2013}. However, this simple description provides a first order approach with a single free parameter, as shown below. Invoking Stokes equations in the lubrication approximation for the surface layer, and assuming no slip between the glassy and surface layers and no shear at the free surface, leads to: 
\begin{equation}
\label{fullgtfe}
\partial_th+\frac{\gamma{h_{\textrm{m}} }^3}{3\eta_{\textrm{m}} }\partial_x^{\,4}h=0\ ,
\end{equation}
which we will refer to as the glassy thin film equation (GTFE). It is mathematically identical to the linearized TFE. The GTFE thus has an exact analytical solution for a stepped initial profile~\cite{Salez2012b} which is self-similar in the variable $xt^{-1/4}$. This solution can be used to extract a single free parameter describing the flow: $\gamma h_{\textrm{m}} ^{3}/(3\eta_{\textrm{m}})$. The form of Eq.~(\ref{fullgtfe}) is mathematically identical to the Mullins model~\cite{Mullins} describing the evolution of profiles by surface diffusion of molecules. However, for flow of macromolecules where all the segments must move together, the GTFE interpretation in terms of surface flow in a layer of size $h_{\textrm{m}}$ is more relevant than this collective surface diffusion scenario. Figure~1 displays a schematic diagram of the two flow regimes studied. The first is for $T>T_{\textrm{g}}$ with homogeneous viscosity (TFE), and the second for $T<T_{\textrm{g}}$ where there is only a thin layer of mobile fluid atop an immobile glassy film (GTFE). The self-similar nature of both Eqs.~(\ref{fulltfe}) and (\ref{fullgtfe}) implies that by fitting their solutions to the experimental profiles one can determine the physical quantities of the problem through a single free parameter. This method of investigation can be carried out with films thick enough that chain confinement and polymer-substrate effects can be ignored. 
\begin{figure}[t!] 
\includegraphics[width=1\columnwidth]{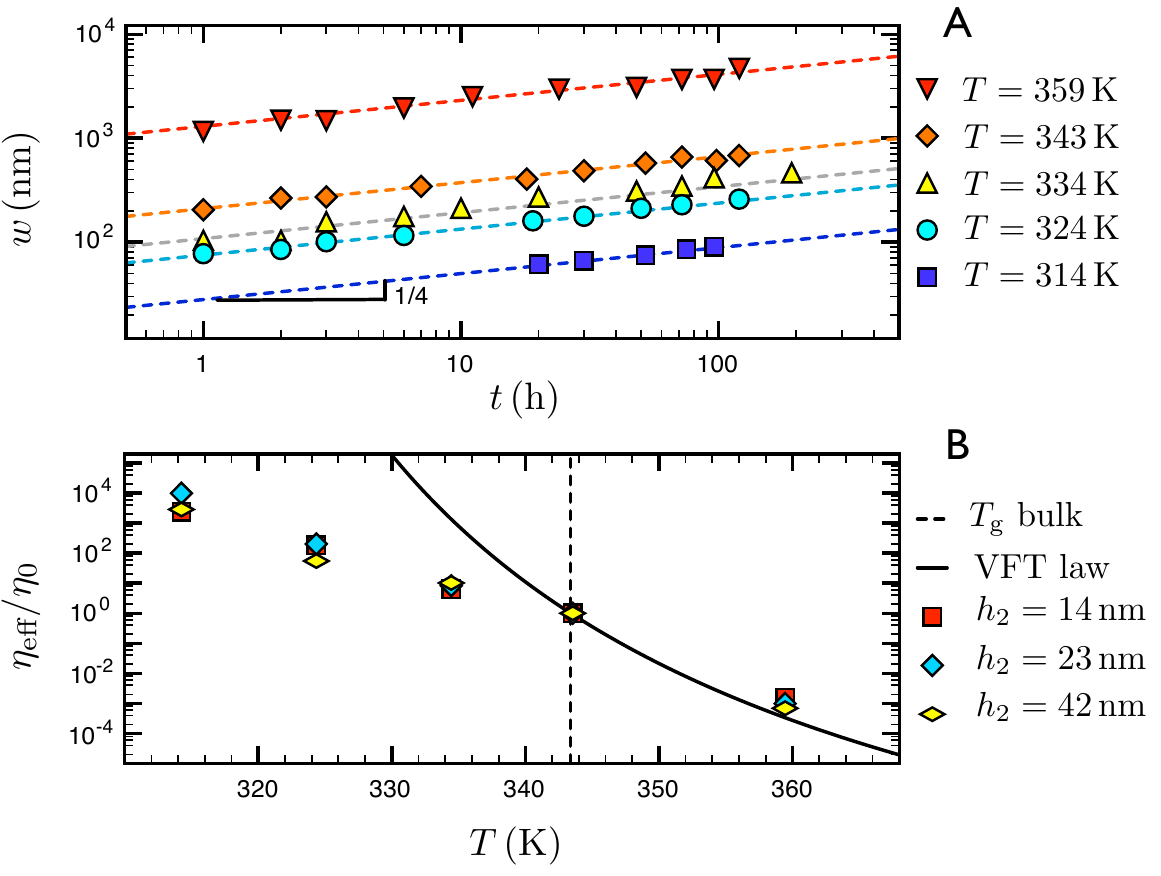}
\caption{Fig. 2. Temporal evolution of the width of stepped films and temperature dependence of the effective viscosity. \textbf{(A)} The temporal evolution of the width $w$ (see Fig.~1C), obtained by fitting the profile to a $\tanh(x/w)$ function, for $h_1=90$~nm, $h_2=42$~nm, at temperatures near the bulk $T_{\textrm{g}}=343~\textrm{K}$. The solid lines have slope $1/4$. \textbf{(B)} Effective viscosities (see definition in text) normalized to the one at $T_0=T_\textrm{g}$ for a given geometry, for films with $h_1=90$~nm and $h_2$ as indicated. Errors are comparable to the symbol size.}
\label{fig2}
\end{figure}

Films were prepared by spin-coating from a dilute solution of polystyrene (PS) dissolved in toluene onto two types of substrates: silicon (Si) with the native oxide layer, and freshly cleaved mica substrates. The PS had weight averaged molecular weight $M_\textrm{w}=3.0\ \textrm {kg.mol}^{-1}$, and polydispersity index 1.09 (Polymer Source Inc.). Samples were annealed at $348$~K, which is $5$~K above bulk $T_{\textrm{g}}$, in an oven flushed with dry N$_2$ for 12 hours. Films with thickness $h_2$ on mica substrates were floated onto the surface of purified water in order to separate the films from the mica. The previously coated Si substrates, coated with PS of thickness $h_1$, were then dipped into the water and used to pick up the floating films. These low molecular weight films are fragile when floating on the water surface and break into smaller sections with several straight vertical edges. Thus, when transferred, these `float-gaps' form perfect steps of height $h_2$ over bottom films of height $h_1$ (see Fig.~1A). The dilatometric $T_{\textrm{g}}$ of independent, annealed flat films on Si was measured by ellipsometry. For $h \geq 40$ nm, we found $T_{\textrm{g}}=343 \pm 2$~K. 

We first demonstrate that the stepped film samples do level at temperatures much less than the bulk $T_{\textrm{g}}$. We performed a simple width evolution experiment where three types of stepped polymer films were prepared with the same bottom layer thickness, $h_1=90$~nm, and top layer thicknesses of $h_2= 14$, 23, and 42~nm (see Fig.~1A). The stepped films were collectively heated in a N$_2$ filled oven and removed after various annealing times for measurement at room temperature with atomic force microscopy (AFM, JPK Instruments). Scan lines were averaged to produce a profile, and the width $w$ (see Fig.~1C) was obtained by fitting this profile to a $\tanh(x/w)$ function. Figure~2A shows the temporal evolution of $w$ for this series of stepped films at five different temperatures. There is an increase with time of the width at temperatures as much as 30~K below the bulk $T_{\textrm{g}}$ value. This indicates enhanced mobility in the glassy film. Furthermore, as suggested by both Eqs.~(\ref{fulltfe}) and (\ref{fullgtfe}), the width varies as $w = (at)^{1/4}$ at all temperatures, where $a$ is a factor that depends a priori on temperature and initial geometry. This $1/4$ power law demonstrates the existence of a capillary-driven flow both above and below $T_{\textrm{g}}$. By analogy with the scaling analysis of Stillwagon and Larson~\cite{StillWL}, a simple determination of the effective viscosity $\eta_\textrm{eff}$ of the sample can be obtained by the vertical offset between the lines in Fig.~2A, using $\eta_{\textrm{eff}}\propto a^{-1}$. The effective viscosity corresponds to the viscosity of the sample calculated as if flow occurs in the entire film, within the lubrication approximation. For a given geometry, it is then possible to compare the $\eta_\textrm{eff}$ values obtained at all temperatures to one, $\eta_0$, at a particular reference temperature $T_0$, in order to get a relative measure of the effective viscosity of the entire film. Setting $T_0 =  T_\textrm{g}$, Fig.~2B shows the relative effective viscosity $\eta_\textrm{eff}/\eta_0 = a_0/a$, for all film geometries. The solid line in this plot is the Vogel-Fulcher-Tammann (VFT) law for PS~\cite{VFT}. When $T>T_{\textrm{g}}$, the temperature dependence of the effective viscosity agrees quantitatively with the bulk VFT law. However, for $T<T_{\textrm{g}}$ there is significant deviation away from this line. There are two ways in which this difference can be interpreted: either i) the entire film flows with viscosity reduced below that predicted by the VFT law (either because the viscosity is reduced from the bulk one, or because the bulk viscosity deviates from the VFT law for $T<T_{\textrm{g}}$), or ii) the assumption of whole film flow is invalid.

In order to distinguish between these two scenarios, we turn to a more quantitative investigation based on the TFE and GTFE. Stepped polymer films with $h_1=h_2=90$~nm were used. Typically, AFM images were collected over a square region of size $\sim 3 \ \mu \textrm{m} \times 3 \ \mu \textrm{m}$ for glassy samples and up to $\sim 50  \ \mu \textrm{m}  \times 50  \ \mu \textrm{m}$ for melt samples. This measurement was repeated until the shape of the profile was self-similar in the variable $xt^{-1/4}$ or, for cases of the two temperatures in the transition region, until a sample was heated for a total of 90 hours. All AFM measurements were carried out at room temperature. For the TFE (see Eq.~(\ref{fulltfe})) and GTFE (see Eq.~(\ref{fullgtfe})), the long-time solutions have been shown to be self-similar in the variable $xt^{-1/4}$~\cite{McGrawPRL,Salez2012a,Salez2012b}. Therefore, if we plot the film height $h(x,t)$ as a function of $xt^{-1/4}$ for several times, the profiles should superimpose. 
\begin{figure}[t!] 
\includegraphics[width=1\columnwidth]{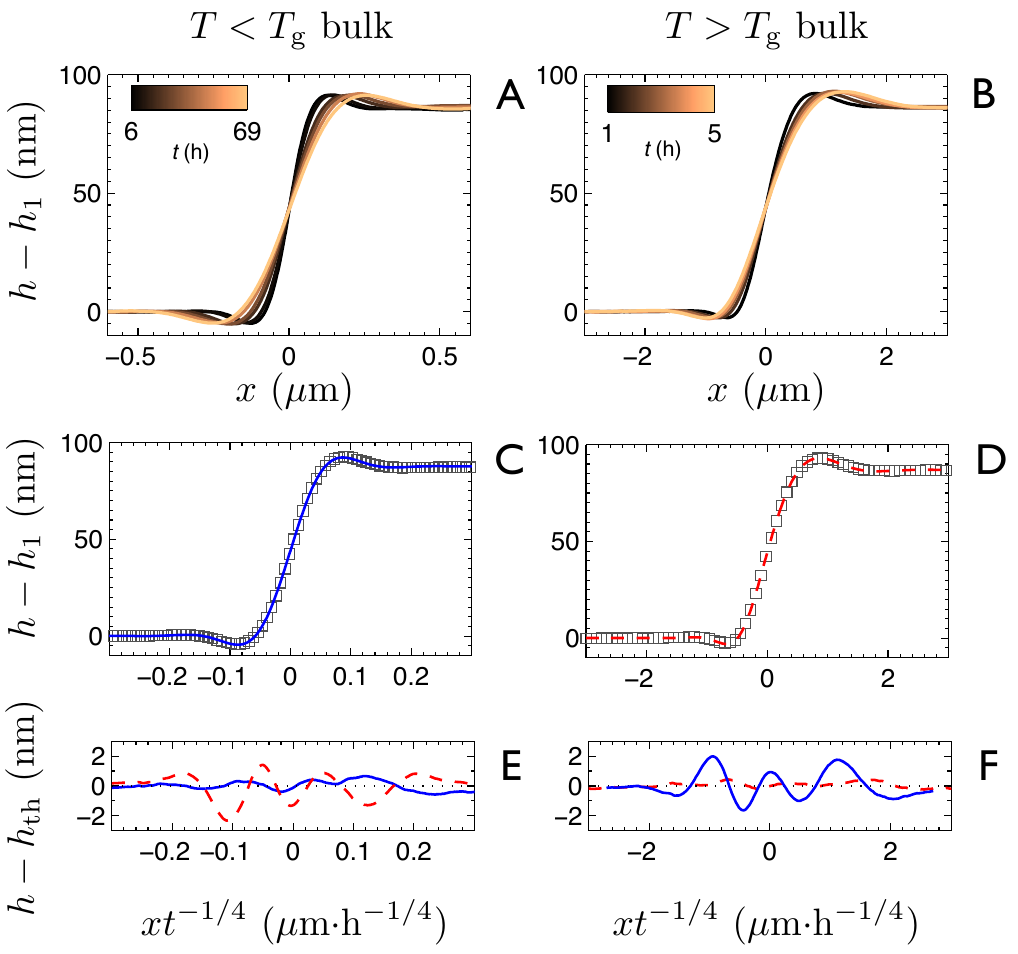}
\caption{Fig. 3. Height profiles and fits below and above $T_{\textrm{g}}$. In both the glassy (left, $T=333$~K) and melt (right, $T=353$~K) cases, the top \textbf{(A} and \textbf{B)} show the temporal evolutions of experimental profiles with $h_1=h_2=90$~nm; the centre \textbf{(C} and \textbf{D)} are the collapsed experimental profiles (white squares) showing self-similar behaviour in the variable $xt^{-1/4}$; and the bottom \textbf{(E} and \textbf{F)} demonstrates the goodness of fits of each collapsed profile to either the TFE numerical solution or the GTFE analytical solution. In \textbf{(C, D, E} and \textbf{F)}, the blue solid line corresponds to the GTFE and the red dashed line corresponds to the TFE.}
\label{fig3}
\end{figure}
Figures~3A and 3B show a number of measured profiles over a large time window, both for temperatures below $T_{\textrm{g}}$ (left) and above $T_{\textrm{g}}$ (right). Figures~3C and 3D show that the profiles are indeed self-similar. While the data obeys this self-similarity for $T<T_{\textrm{g}}$ and $T>T_{\textrm{g}}$, there are important differences between the two temperature regimes. In particular, the shapes of the self-similar profiles are different. See for example Figs.~3C and 3D, where one can see that above $T_{\textrm{g}}$ the magnitude of the bump (first top oscillation) is larger than that of the dip (first bottom oscillation). Below $T_{\textrm{g}}$, it is similarly evident that the bump and dip extrema are equal in magnitude. To be more precise, above $T_{\textrm{g}}$ these features depend quantitatively on $h_1$ and $h_2$, whereas below $T_{\textrm{g}}$ the surface flows without sensitivity to the substrate for the considered thicknesses. In the latter case, samples with same $h_2$ but different $h_1$ show bumps and dips that are all equal in magnitude. This simple qualitative feature of the profiles shows that it is the surface alone that flows below $T_{\textrm{g}}$. Fits of the sub-$T_{\textrm{g}}$ profiles to solutions of both the TFE and GTFE quantitatively highlight the differences. The left plots of Fig.~3 are for $T<T_{\textrm{g}}$. The blue solid line in Fig.~3C is a best fit of the self-similar experimental profile to the GTFE analytical solution~\cite{Salez2012b}. The residuals of the fit are also shown as a blue solid line in Fig.~3E. The red dashed line in Fig.~3E corresponds to the fit of the sub-$T_{\textrm{g}}$ data to the TFE numerical solution~\cite{Salez2012a}. In the sub-$T_{\textrm{g}}$ case, the experimental profiles are thus much better described by the GTFE than by the TFE, as the residuals are much lower and do not exhibit the systematic variation seen in the dashed residuals. Similarly, the experimental data on the right side of Fig.~3 are for $T>T_{\textrm{g}}$ and are much better represented by the TFE than by the GTFE.   

Since the TFE and GTFE correspond to different physical pictures, we can define a single metric, $\chi$, for the goodness of fit to each model.  $\chi$ is used in order to characterize the transition from where the system is best described by whole film flow, to where it is best described by surface flow over thickness $h_{\textrm{m}}$. We define this quantity by the correlation function:
\begin{equation}
\label{eqcor}
\chi = \frac{\int dx\, (h_{\textrm{EXP}} - h_\textrm{TFE})^2}{\int dx\, (h_\textrm{GTFE} - h_\textrm{TFE})^2}\ , 
\end{equation}
where $h_\textrm{EXP}$ is the self-similar experimental profile, $h_\textrm{TFE}$ is the numerical solution~\cite{Salez2012a} of the TFE, and $h_\textrm{GTFE}$ is the analytical solution~\cite{Salez2012b} of the GTFE. This function equals 1 if the experimental data is exactly described by the GTFE solution, and 0 if the experimental data is exactly described by the TFE solution. Figure~4 shows the temperature dependence of $\chi$ as well as the temperature dependence of the thermal expansivity $\frac{dh}{dT}$ derived from ellipsometry measurements for an independent flat $87$~nm thick film. This type of ellipsometry data is often used to find the dilatometric $T_{\textrm{g}}$ value in thin films~\cite{JAFKDVreview}, and in this case gives rise to  $T_{\textrm{g}}=343 \pm 2$~K. It is remarkable that $\chi(T)$ undergoes an abrupt transition at a temperature indistinguishable from the bulk $T_{\textrm{g}}$ value. This means that the system undergoes a sharp transition from bulk flow to surface dominated flow as the temperature is lowered through the bulk $T_{\textrm{g}}$ value. The transition temperature should be interpreted as the one below which most of the film exhibits no flow on the 90 hours time scale. 

Polystyrene is a model glass-forming material, and through our measurements, we should be able to probe other aspects of the glassy dynamics. In particular, while the data shown in Fig.~4 is for profiles having reached a steady-state value of $\chi$ or after 90 hours of annealing, whichever occurs  first, we can measure the time dependence of the shape of the profile at temperatures near the transition. For temperatures in this range, we might expect to see evidence of the time dependent mechanical properties of the glassy material. In particular, glass-forming materials behave like elastic solids on short time scales, and like viscous materials on much larger time scales. We thus might expect that for short times the system would behave like a glassy material, with flow only occurring in the surface region and the profile well described by the GTFE; and for long times  the system would be well characterized by flow of the entire film with a profile well described by the TFE. The inset shows the temporal evolution of $\chi$ for the particular $343$~K and $348$~K data, which lie in the transition region. For the case of  $T=343$~K, the initial $\chi$ value is near 1, meaning the system is initially dominated by flow localized in a surface region. As the system evolves in time, this correlation decreases, and the system becomes less well described by the GTFE. The sample at $T=348$~K shows even more striking behaviour. In this case, one can see that over a period of $3 \times 10^5$~s the system goes from being well described by flow localized in the surface region ($\chi \sim 1$), to being well described by flow in the entire film ($\chi \sim 0$). In this transition region, because the shape of the profile is changing from glass-like to fluid-like, the profiles cannot be self-similar over a large time window. 
\begin{figure}[t!] 
\includegraphics[width=1\columnwidth]{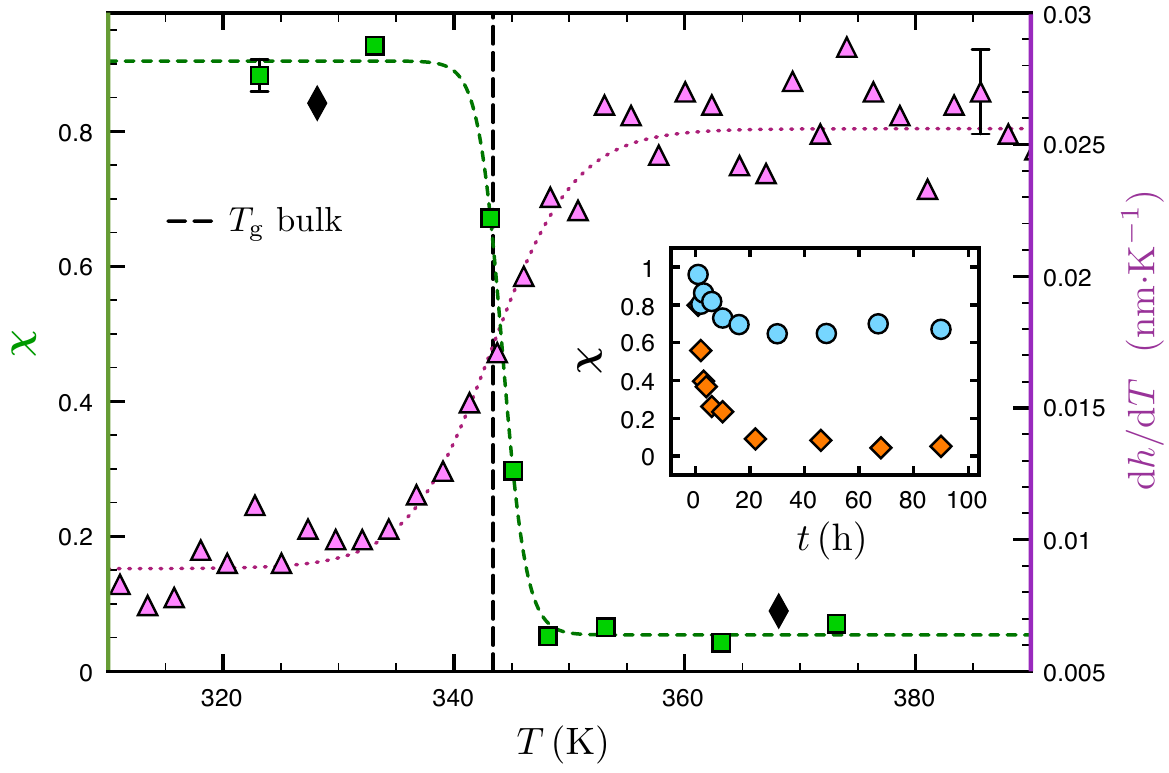}
\caption{Fig. 4. Temperature dependence of the correlation function and thermal expansivity. The correlation function $\chi(T)$ defined in Eq.~(\ref{eqcor}) is given by the green square and black diamond symbols (left axis) for samples with $h_1=h_2=90$~nm. The thermal expansivity for an independent flat $87$~nm sample is given by the purple triangles  (right axis). The black diamond symbols are $\chi(T)$ for a single sample that was held first for 90 hours at $T< T_{\textrm{g}}$, then measured and heated to $T> T_{\textrm{g}}$ until the self-similar profile was reached. The inset shows the temporal evolution of $\chi$ for $T=343$~K and $348$~K data that lie in the transition region (blue circles are for $T= 343$~K, orange diamonds are for $T=348$~K). Error bars are indicated once for each subplot.}
\label{fig4}
\end{figure}
 
The time dependences of $\chi$ at $T = 343$ and 348\ K show that the height profiles can be used to monitor the system as it changes from glassy-like to liquid-like behaviour. This also implies that we could probe in a single sample more than one temperature, as long as we waited for the profile to reach a self-similar state at each temperature. The data shown with diamond symbols in Fig.~4 show  that a single sample, with a film thickness large enough so that chain confinement and substrate effects can safely be neglected, can exhibit a transition from bulk flow to surface flow. This thickness range can also be used for molecular glasses where other experiments requiring thinner films would not be appropriate, because dewetting is too rapid above $T_{\textrm{g}}$. 

The numerical fits to the data can be used to extract meaningful physical parameters. In both the TFE and GTFE cases, there is a single free horizontal stretching parameter that determines the fit of the self-similar experimental profile to the dimensionless theoretical solution. Knowing the tabulated~\cite{surfacetension} surface tension $\gamma$, the GTFE fitting parameter (see Eq.~(\ref{fullgtfe})) gives the surface mobility $h_{\textrm{m}} ^{3}/(3\eta_{\textrm{m}})$, and the TFE fitting parameter (see Eq.~(\ref{fulltfe})) gives the bulk viscosity $\eta_{\textrm{b}}$. As a more direct comparison, we plot the GTFE mobility $h_{\textrm{m}} ^{3}/(3\eta_{\textrm{m}})$ and the average TFE mobility $(h_1+\frac{h_2}{2})^3/(3\eta_{\textrm{b}})$ on the single composite plot of Fig.~5. The result is consistent with that of Yang \textit{et al.} \cite{tsuiscience} but with two differences in the methodology: we use films that are thick enough to prevent chain confinement and substrate effects, and we have the possibility of using a single sample to obtain the entire curve. The solid line is obtained from using the bulk VFT law for PS of the same $M_\textrm{w}$~\cite{VFT}. The agreement we obtain for $T>T_{\textrm{g}}$ (left side of Fig.~5) is consistent with the previous success of the stepped film technique in polymer melts~\cite{McGrawPRL}. Of more importance for the present work is the sub-$T_{\textrm{g}}$ mobility (right side of Fig.~5), for which we observe a strong deviation from the bulk VFT law. In this temperature range, the single fit parameter $h_{\textrm{m}} ^{3}/(3\eta_{\textrm{m}})$ combines the two relevant physical quantities of the mobile surface layer: its size and viscosity. Using reasonable constraints we now estimate each parameter individually. In order for  any flow to occur, the size of the surface region has to be large enough so that the polymer molecules fit into it. Molecules larger than the surface region size will have segments in the glassy region, and will thus be unable to flow. For PS with $M_\textrm{w}=3\ \textrm {kg.mol}^{-1}$, the root-mean-squared end-to-end distance of the molecule satisfies $\langle R_{\textrm{EE}} \rangle_{\textrm{RMS}} \sim 3$~nm, and we can use this as a first estimate for the surface region size that is similar to the one of Refs.~\cite{PRE2000,PRERC2000}. This length scale coupled with the data in Fig.~5 suggests a surface viscosity of $\eta_{\textrm{m}}  \sim 2\times10^8  \textrm{ Pa}\cdot\textrm{s}$ at $323$~K. We note that we used only an estimate for $h_{\textrm{m}} $ and the subsequent estimated value of $\eta_{\textrm{m}}$ is very sensitive to this chosen thickness. In particular, the constraint $h_{\textrm{m}}  > \langle R_{\textrm{EE}} \rangle_{\textrm{RMS}}$ may be too strong, as this is an average of the molecular size, and it is only necessary that there is a significant fraction of the molecules that have all segments in the surface region in order to have surface flow. Alternatively, if we used the value of $h_{\textrm{m}} \sim 1$~nm suggested as a lower limit in ref \cite{ChadTNB}, we would predict a surface viscosity of $7.7\times 10^6 \textrm{ Pa}\cdot\textrm{s}$ at $323$~K, which is $20$~K below $T_{\textrm{g}}$. This surface viscosity is more than 3 orders of magnitude lower than the bulk viscosity at $T_{\textrm{g}}$. Finally, the observed linear trend (in log-lin scale) of Fig.~5 below $T_{\textrm{g}}$ allows us to infer an Arrhenius behaviour of the surface mobility below $T_{\textrm{g}}$, with activation energy $E_{\textrm{a}}\approx337\pm20\ \textrm{kJ}.\textrm{mol}^{-1}$ in agreement with existing literature \cite{zahrascience,Papaleo,tsuiscience}. 
\begin{figure}[t!] 
\includegraphics[width=1\columnwidth]{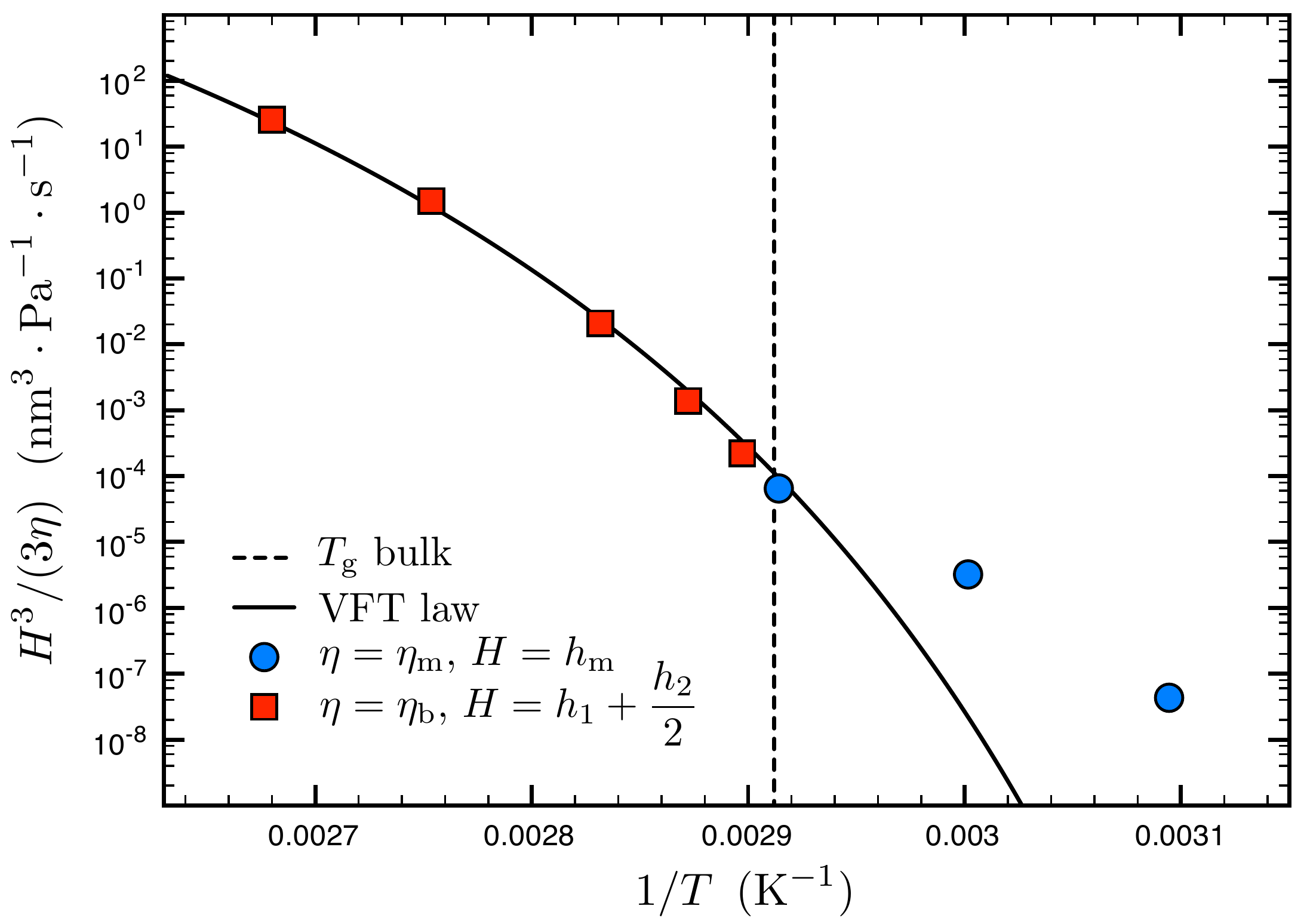}
\caption{Fig. 5. Temperature dependence of the mobility. This figure is made of two subplots representing $h_1=h_2=90$~nm samples. The mobility $H^3/(3\eta)$ is determined by a fit to either the analytical GTFE solution (blue circles) or the numerical TFE solution (red squares), with $\eta$ and $H$ as defined in the legend.}
\label{fig5}
\end{figure}

In conclusion, by employing the stepped film geometry and analysing the resulting flow, we report quantitative evidence for the existence of a thin layer of liquid-like material at the free surface of glassy, low molecular weight polystyrene films. The sample thicknesses and preparation are such that annealing effects, chain confinement, and substrate effects can be neglected. The transition from whole film flow to flow localized in a thin surface layer has been measured and observed to occur sharply at the bulk $T_{\textrm{g}}$ value. For temperatures inside the transition region, we were able to measure time dependent evolutions from glassy to liquid behaviour. This technique provides an opportunity to accurately follow the transition from surface flow to bulk flow within a single sample. Below $T_{\textrm{g}}$, a fit to the measured profile gives a surface mobility parameter $h_{\textrm{m}} ^{3}/(3\eta_{\textrm{m}})$ that can be used to estimate a surface viscosity. In particular, we obtain $\eta_{\textrm{m}} \sim 10^8 \textrm{~Pa} \cdot \textrm{s}$ at $20$~K below $T_{\textrm{g}}$. Independent determination of either the size $h_{\textrm{m}} (T)$ of the surface region or its viscosity  $\eta_{\textrm{m}} (T)$ would allow a complete determination of the temperature dependent properties of the near surface region. 

\section*{Acknowledgements}
The authors would like to acknowledge the \'Ecole Normale Sup\'erieure of Paris, the Fondation Langlois, the Chaire Total - ESPCI ParisTech, as well as the NSERC of Canada for financial support. They also thank Oliver B\"aumchen for interesting discussions.


\begin{thebibliography}{}
\bibitem{edigerreview} M. D. Ediger, P. Harrowell, J. Chem. Phys. {\bf 137}, 080901 (2012).
\bibitem{andersonscience} P. W. Anderson, Science {\bf 267}, 1615 (1995).
\bibitem{EPJEreview} J. A. Forrest, Eur. Phys. J. E {\bf 8}, 261 (2002).
\bibitem{JAFKDVreview} J. A. Forrest, K. Dalnoki-Veress, J. Coll. Int. Sci. \textbf{94/1-3}, 167 (2001).
\bibitem{McKennareview} M. Alcoutlabi and G. B. McKenna, J. Phys.: Condens. Matter \textbf{17}, R461 (2005).
\bibitem{teisseire11APL}J. Teisseire, A. Revaux, M. Foresti and E. Barthel, App. Phys. Lett. \textbf{98}, 013106 (2011).
\bibitem{rognin11PRE}E. Rognin, S. Landis, and L. Davoust, Phys. Rev. E \textbf{84}, 041805 (2011).
\bibitem{rognin12JVS} E. Rognin, S. Landis, and L. Davoust, J. Vac. Sci. Technol. B  \textbf{30}, 011602 (2012).
\bibitem{contradict1} T. Kerle, Z. Lin, H. Kim, and T. P. Russell, Macromolecules {\bf 34}, 3484 (2001).
\bibitem{contradict2} S. Ge \textit{ et al.}  Phys. Rev. Lett. {\bf 85}, 2340 (2000).
\bibitem{zhu11PRL}L. Zhu, C. Brian, S. Swallen, P. Straus, M. D. Ediger, and L. Yu, Phys. Rev. Lett.  \textbf{106}, 256103 (2011).
\bibitem{edigerjcp} R. Malshe, M. D. Ediger, L. Yu, J. J. de Pablo, J. Chem. Phys. {\bf 134}, 194704 (2011).
\bibitem{embeddingprl}J. H. Teichroeb, J. A. Forrest, Phys. Rev. Lett. {\bf 91}, 016104 (2003).
\bibitem{embeddingilton}M. Ilton, D. Qi, J. A. Forrest, Macromolecules {\bf 42}, 6851 (2009).
\bibitem{embeddingQi} D. Qi, M. Ilton, J. A. Forrest, Eur. Phys. J. E. {\bf 34}, 56 (2011).
\bibitem{embeddingMw} D. Qi, Ph.D. thesis, University of Waterloo, 2009.
\bibitem{ChadTNB}C. R. Daley, Z. Fakhraai, M. D. Ediger, and J. A. Forrest, Soft Matter {\bf 8} 2206 (2012).
\bibitem{zahrascience}Z. Fakhraai, J. A. Forrest, Science  {\bf 319}, 600 (2008).
\bibitem{QiPRL}D. Qi, Z. Fakhraai, and J. A. Forrest, Phys. Rev. Lett. {\bf 101}, 096101 (2008).
\bibitem{Papaleo} R. M. Papal\'{e}o, R. Leal, W. H. Carreira, L. G. Barbosa, I. Bello, A. Bulla, Phys. Rev. B {\bf 74}, 094203 (2006).
\bibitem{buck04MAC}E. Buck, K. Petersen, M. Hund, G. Krausch and D. Johannsmann, Macromolecules \textbf{37}, 8647 (2004).
\bibitem{Mullins} W. W. Mullins, J. Chem. Phys. {\bf 30}, 77 (1959).
\bibitem{tsuiscience} Z. Yang, Y. Fujii, F. K. Lee, C.-H. Lam, O. K. C. Tsui, Science {\bf 328}, 1676 (2010).
\bibitem{McGrawPRL} J. D. McGraw, T. Salez, O. B\"{a}umchen, E. Rapha\"{e}l and K. Dalnoki-Veress, Phys. Rev. Lett. \textbf{109}, 128303 (2012).
\bibitem{Salez2012a} T. Salez, J. D. McGraw, S. L. Cormier, O. B\"{a}umchen, K. Dalnoki-Veress and E. Rapha\"{e}l, Eur.  Phys. J.  E \textbf{35}, 114 (2012).
\bibitem{Salez2012b} T. Salez, J. D. McGraw, O. B\"{a}umchen, K. Dalnoki-Veress and E. Rapha\"{e}l, Phys. Fluids \textbf{24}, 102111 (2012).
\bibitem{Blossey} R. Blossey, {\em Thin liquid films: dewetting and polymer flow} (Springer, Dordrecht, 2012).
\bibitem{Forrest2013}J. A. Forrest,  J. Chem. Phys. {\bf 139}, 084702 (2013).
\bibitem{StillWL} L. E. Stillwagon, R.G. Larson,  J. Appl. Phys. {\bf 63}, 5251 (1988).
\bibitem{VFT} J.-C. Majeste, J.-P. Montfort, A. Allal, G. Marin, Rheol. Acta {\bf 37} 486 (1998).
\bibitem{surfacetension} S. Wu, J. Chem. Phys. {\bf 74}, 632 (1970).
\bibitem{PRE2000} J. Mattsson , J. A. Forrest, L. Borjesson, Phys. Rev. E {\bf 62}, 5187, (2000).
\bibitem{PRERC2000}J. A. Forrest, J. Mattsson, Phys. Rev. E {\bf 61} R53 (2000).
\end{thebibliography}
\end{document}


\title{\ A direct quantitative measure of surface mobility in a glassy polymer (Supplementary Materials)}
\author{Y. Chai}
\affiliation{Department of Physics \& Astronomy and Guelph-Waterloo Physics Institute, University of Waterloo,  Waterloo, Ontario, Canada, N2L 3G1.}
\author{T. Salez}
\affiliation{Laboratoire de Physico-Chimie Th\'eorique, UMR CNRS Gulliver 7083, ESPCI ParisTech, Paris, France}
\author{J. D. McGraw}
\altaffiliation{Department of Experimental Physics, Saarland University, 66041 Saarbr\"{u}cken, Germany}
\affiliation{Department of Physics and Astronomy, McMaster University,  Hamilton, Ontario, Canada, L8S 4M1.}
\author{M. Benzaquen}
\affiliation{Laboratoire de Physico-Chimie Th\'eorique, UMR CNRS Gulliver 7083, ESPCI ParisTech, Paris, France}
\author{K. Dalnoki-Veress}
\affiliation{Laboratoire de Physico-Chimie Th\'eorique, UMR CNRS Gulliver 7083, ESPCI ParisTech, Paris, France}
\affiliation{Department of Physics and Astronomy, McMaster University,  Hamilton, Ontario, Canada, L8S 4M1.}
\author{E. Rapha\"{e}l}
\affiliation{Laboratoire de Physico-Chimie Th\'eorique, UMR CNRS Gulliver 7083, ESPCI ParisTech, Paris, France}
\author{J. A. Forrest}
\thanks{corresponding author. email: jforrest@uwaterloo.ca}
\affiliation{Department of Physics \& Astronomy and Guelph-Waterloo Physics Institute, University of Waterloo,  Waterloo, Ontario, Canada, N2L 3G1.}
\date{\today}
\maketitle

\section{Influence of the film thickness}
\begin{figure}[h!]
\begin{center}
\begin{minipage}[t]{7.2cm}
\centering
\includegraphics[width=7.2cm]{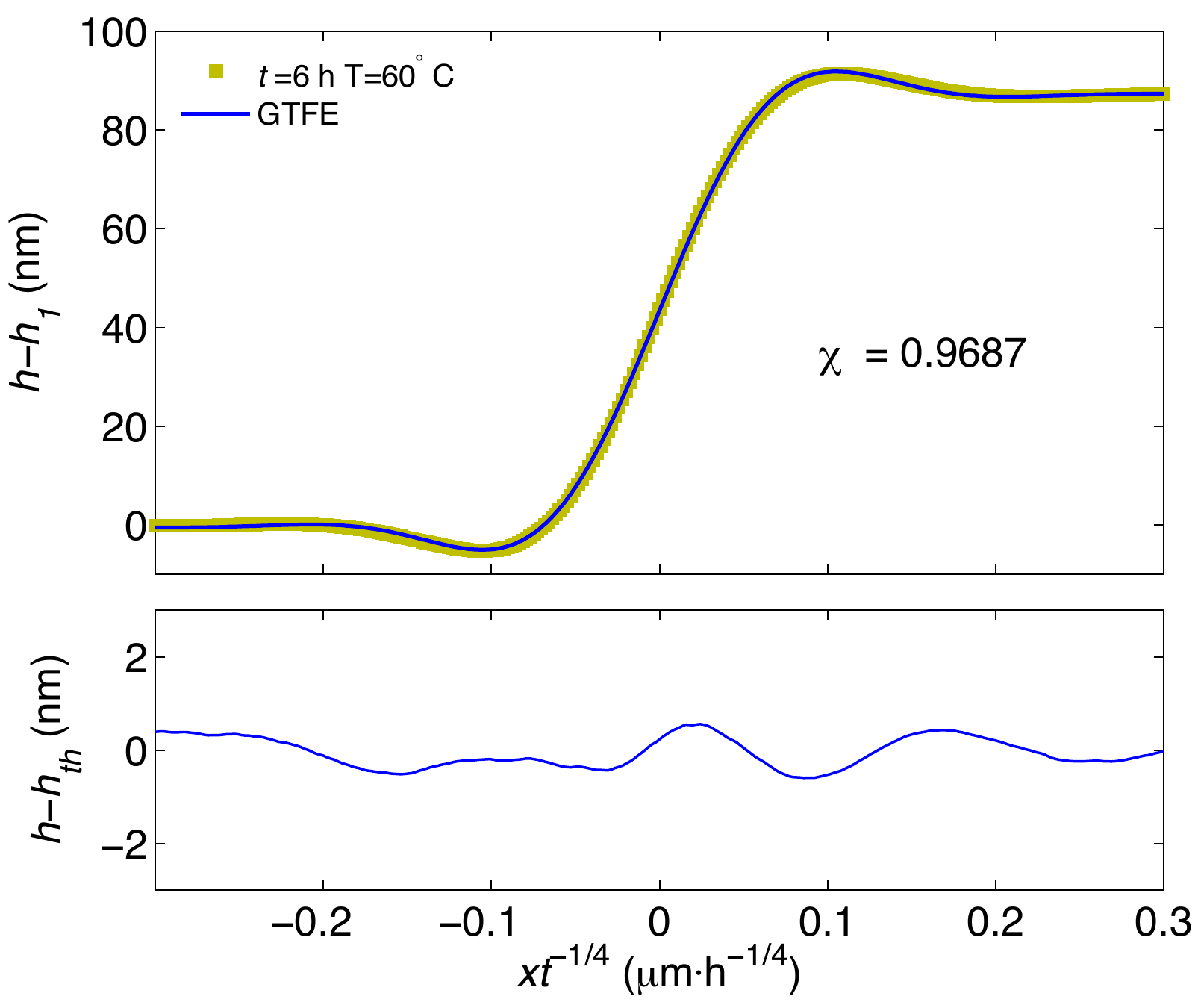}
\end{minipage}
\begin{minipage}[t]{7.2cm}
\centering
\includegraphics[width=7.2cm]{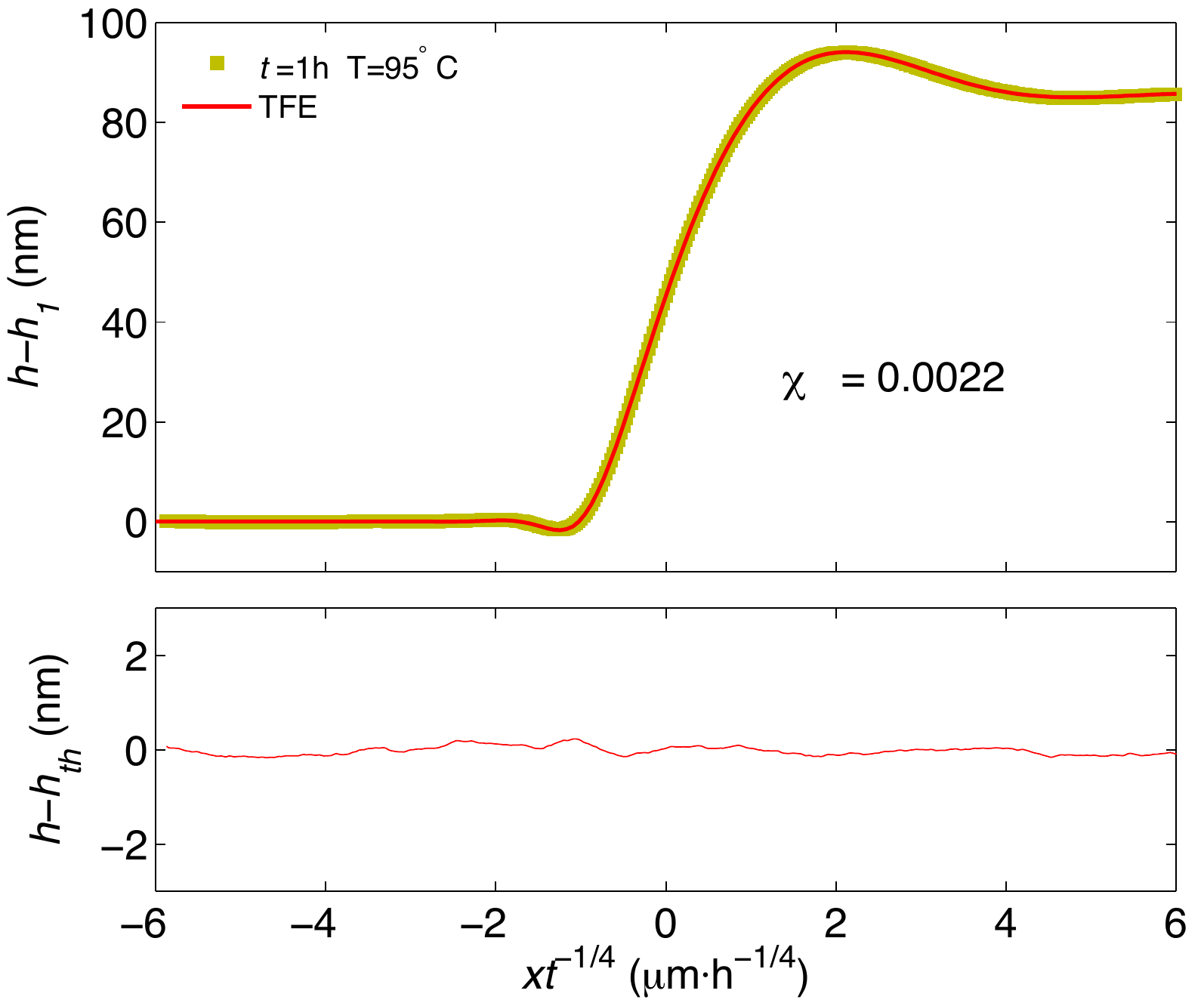}
\end{minipage}
\caption{\textit{Profiles for stepped films with  $\{h_1, h_2\} = \{23, 90\}$\ nm.
The shifted experimental profiles, $h\left(xt^{-1/4}\right)-h_1$, are shown at two different temperatures $T$. The solid lines are the fits to the corresponding theoretical profiles, $h_{\textrm{th}}\left(xt^{-1/4}\right)-h_1$, which are the GTFE analytical solution below $T_{\textrm{g}}$ (top, see Eq.~(2)) and the TFE numerical solution above $T_{\textrm{g}}$ (bottom, see Eq.~(1)), together with the corresponding residual plots. The values of the correlation function, $\chi$, introduced in the article at Eq.~(3) are also indicated.}}
\label{figS1}
\end{center}
\end{figure}

A sample configuration with $\{h_1, h_2\} = \{23, 90\}$\ nm was also studied. The corresponding fits, at two different temperatures, are shown in Fig.~\ref{figS1}. Below $T_{\textrm{g}}$ ($T=T_{\textrm{g}}-10$~K), the profile is best fit to the GTFE (see Eq.~(2)) analytical solution. The GTFE dimensionless profile $(h_{\textrm{th}}-h_1)/h_2$, with the characteristic bump and dip of equal size, is independent of $h_1$ and $h_2$ due to the linearity of the GTFE. The mobility obtained from this fit is within a factor of 3 of that for samples with $h_1=h_2=90$~nm at the same temperature (see Fig.~5). The difference is comparable to the symbol size in Fig.~5 and indicates that the obtained surface mobility is independent of $h_1$, at this level of description. To sum up, the film thickness does not control the shape of the profile below $T_{\textrm{g}}$. In contrast, this statement is not valid above $T_{\textrm{g}}$ ($T=T_{\textrm{g}}+25$~K). At this temperature, the profile is best fit to the TFE (see Eq.~(1)) numerical solution, for which there is a clear asymmetry in the bump-to-dip aspect ratio. This profile is qualitatively different from the TFE fit reported in Fig.~3, for the $h_1=h_2=90$~\textrm{nm} geometry. The geometrical dependence of height profiles above $T_\textrm{g}$ is described in detail in~\cite{McGraw2012}. 

\section{Validity of the lubrication approximation}
After a sufficient time height profiles will be smooth enough so that the lubrication approximation is valid. One can show~\cite{Salez2012b,Benzaquen2013} that self-similarity is always reached in a finite time for a random initial condition in the TFE and linearised~\footnote{Which is mathematically identical to the GTFE~\cite{Salez2012a}.} TFE. Since the TFE and GTFE are based on the lubrication approximation and their solutions show long-term self-similarity in $xt^{-1/4}$, this long-term self-similarity is an indication of lubrication. Therefore, the criterium we set for an experimental profile to be considered is that self-similarity is reached. For this reason, we only report profiles that have reached self-similarity as shown in Figs.~3C and 3D. Below, we give details on the two main ingredients of the lubrication approximation used to obtain the thin film equations: namely the small slope and small thickness assumptions.

Both the TFE (see Eq.~(1)) and GTFE (see Eq.~(2)) assume that $|dh/dx|\ll1$. It is important to consider whether the  actual measured values of these parameters justify the use of the TFE and GTFE. As shown in Figs.~3A and 3B for $T=333$~K and $T=353$~K, the worst case corresponds to sub-$T_{\textrm{g}}$ experiments at early times where one sees a typical width of $\sim250$~nm for a typical height variation of $\sim90$~nm. This means a maximal slope of $\sim0.35$ which is less than $1$. The natural question is whether such a slope is sufficiently small in comparison to $1$ in order to assume validity of the models. While at small times, the condition may not be valid, the fact that self-similarity is eventually reached is an indication that the condition $|dh/dx|\ll1$ is fulfilled at times long enough where we have the right time scaling. In fact, the slope has to satisfy $|dh/dx|\ll1$ in order to assume that the curvature in the Laplace pressure can be approximated by: 
\begin{equation*}
\kappa=-\frac{h''}{(1+h'^2)^{3/2}}\approx - h''\ ,
\end{equation*}
as explicitly used in both the TFE and GTFE. However, keeping the full non-linear expression of the curvature in the equation above would modify the model. In particular, owing to the denominator in the square braces, it is immediately seen that the key self-similarity used in this article, $h(x,t)=f\left(xt^{-1/4}\right)$, is no longer satisfied in a large slope model. The situation would be similar if one allowed a vertical velocity in the film: the self-similarity would no longer be satisfied. Interestingly, a similar discussion has been developed for contact line lubrication~\cite{Snoeijer2006}. As one sees in this reference, the governing equation has to be modified by a nonconstant factor $F(\theta)$ which prevents overall self-similarity (see Eq.~(14) therein). 

The TFE and GTFE also require that the total film thickness $h$ is much less than the horizontal length scale of the surface profile which scales typically like the width of the step profile $w$ (see Fig.~1). This criterion permits the assumption that the perpendicular component of the polymer flow is negligible or the flow is approximately parallel to the substrate. Otherwise, the full hydrodynamic equations would have to be used. For the data shown in Fig. 2A, one has $\langle h\rangle = 90+42/2 = 111$~nm. Quite a few data points and certainly those taken at $314$~K seem to violate the $h \ll w$ requirement. First, we note that in the glassy case, the limiting lubrication criterium is the slope criterium $|dh/dx|\ll1$ since the typical flow thickness to be considered is $h_\textrm{m}$ (which is of the order of a few nanometers) rather than the total height $h$. Secondly, as explained above, the self-similar behaviour in $w\propto t^{1/4}$ is an indication that the lubrication regime is reached even in the $314$~K worst case where $|dh/dx|\sim1$. Thirdly, one should note that Fig.~2 is obtained from the raw data and is not directly dependent on a quantitative comparison with the TFE or the GTFE. What is performed to produce Fig.~2B is rather to assume a thin film literature-based scaling law of the type:  $w\propto(at)^{1/4}$~\cite{Stillwagon1988}, where we assume $a \sim \eta_\textrm{eff}^{-1}$ as in Stillwagon and Larson's analysis. From this, we estimate an effective viscosity $\eta_\textrm{eff}$ and describe a deviation from the bulk VTF law. It is this deviation that motivates the remainder of the article and the development of the GTFE. The effective viscosity of Fig.~2 is simply the viscosity which would correspond to homogeneous bulk flow in the lubrication approximation. 

\section{AFM and raw data traces}
\begin{figure}[h!]
\begin{center}
\includegraphics[width=0.7\columnwidth]{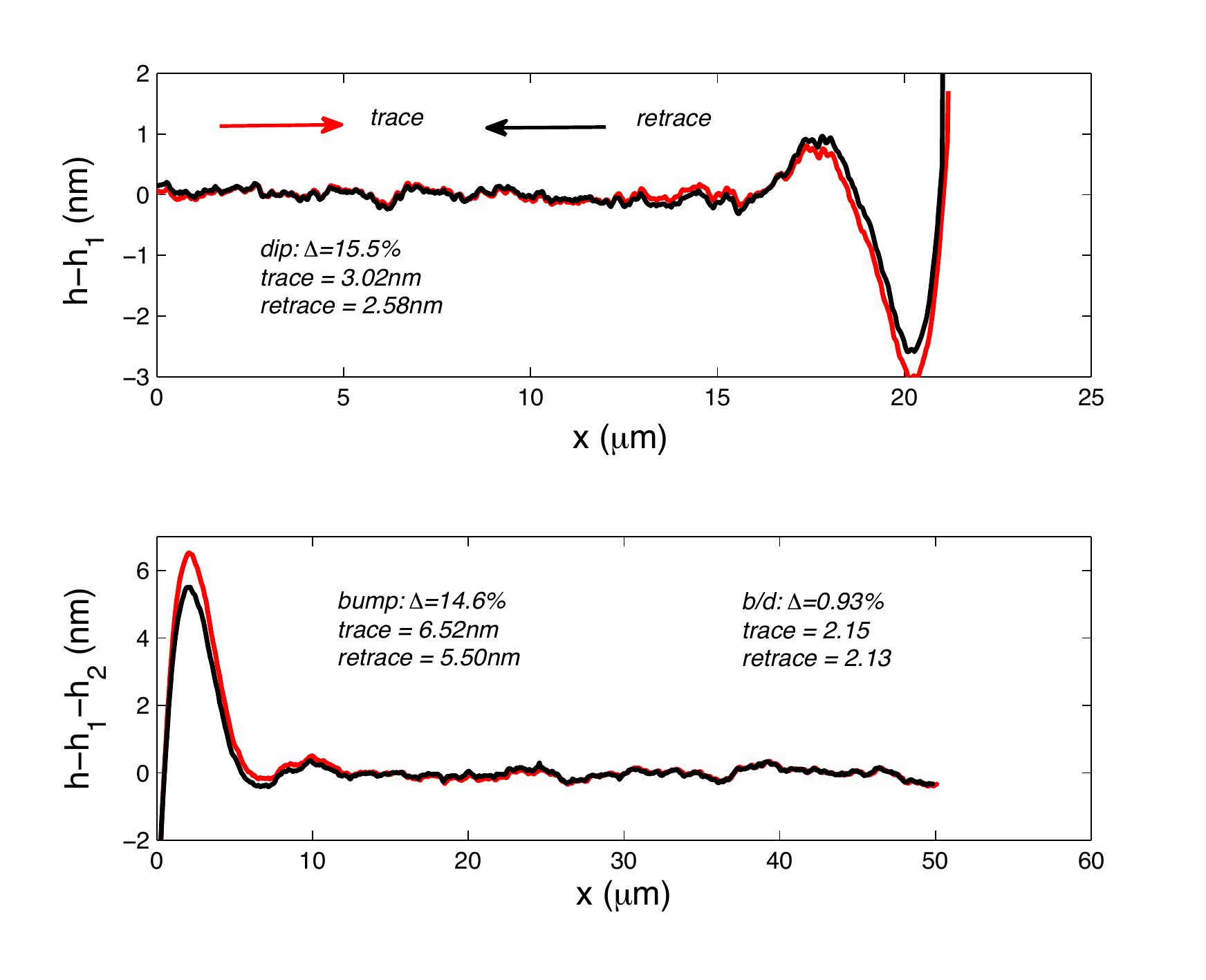}
\end{center}
\caption{\textit{Average of several tens of scan lines around the dip (top) and bump (bottom), from raw AFM ``trace'' and ``retrace" scans.}}
 \label{fig2}
\end{figure}

The goodness of fits (Figs. 3E and 3F) shows that the differences between the TFE and GTFE fits are on the order of 2~nm, which is much less than the step height. It may be a point of  concern  that a main support for the transition between the GTFE and TFE is based on seemingly subtle differences in AFM traces. AFM is known to have high resolution in the vertical direction but not as high in the horizontal direction. A large slope $|dh/dx|$ of the profile may couple the poor horizontal resolution to the vertical resolution through $dh = |dh/dx|dx$. We can easily address this by considering the raw AFM traces shown in Fig.~\ref{fig2}. This figure shows a typical average of a few tens of scan lines. The ``trace" and ``retrace" in the region near the bump and the dip are shown (all of the profiles in the article are similarly averages of many scans). What is clear from this figure is that even the small sub-nanometer details are quantitatively reproduced in terms of the actual height value ($z$) but also in their location ($x$). Thus, the precision in $x$ is more than sufficient to provide the accuracy and resolution required for these measurements. Finally, the transition between the two mobility mechanisms is probed through the integrated quantity $\chi$ (see Eq.~3), which allows to average out the remaining small errors. 
